\documentclass[aps,prab,10pt,reprint,a4paper]{revtex4-2}
\usepackage{graphicx}
\usepackage{amsmath}
\usepackage{amssymb}
\usepackage{multirow}
\usepackage[separate-uncertainty=true]{siunitx}

\newcommand{\E}[1]{\langle #1 \rangle}
\newcommand{\mat}[1]{\underline{#1}}
\newcommand{\T}{\mathrm{T}}
\newcommand{\rmd}{\mathrm{d}}

\newcommand{\R}{\mathbb{R}}
\newcommand{\Z}{\mathbb{Z}}

\newcommand{\A}{\mathcal{A}}
\newcommand{\F}{\mathcal{F}}
\newcommand{\e}{\mathrm{e}}
\newcommand{\I}{\mathrm{i}}
\newcommand{\fref}[2][]{\ref{#2}{#1}}
\newcommand{\figref}[2][]{Fig.~\fref[#1]{#2}}
\newcommand{\Eqq}{\E{q^2}_{\Psi}}

\newcommand{\Eqp}{\E{qp}_{\Psi}}
\newcommand{\Eqqi}{\E{q^2}_{\Psi_i}}
\newcommand{\Eppi}{\E{p^2}_{\Psi_i}}
\newcommand{\Eqpi}{\E{qp}_{\Psi_i}}
\newcommand{\EqqA}{\E{q^2}_{\Psi_A}}

\newcommand{\EqqB}{\E{q^2}_{\Psi_B}}

\newcommand{\Ef}{\mathcal{E}}

\begin{document}

\title{%
Laser-driven bunch compression for ultrashort free-electron laser pulses
}
\date{August 20, 2025}
\author{Ph.~Amstutz}
\email{philipp.amstutz@tu-dortmund.de}
\author{W.~Helml}
\author{S.~Khan}
\author{C.~Mai}
\affiliation{TU Dortmund University,
 Maria-Goeppert-Mayer-Stra\ss{}e 2, 44227 Dortmund, Germany}
\author{Ch.~Gerth}
\author{Ch.~Mahnke}
\author{E.~A.~Schneidmiller}
\affiliation{Deutsches Elektronen-Synchrotron DESY,
 Notkestra\ss{}e 85, 22607 Hamburg, Germany}

\begin{abstract}
Generation of ultrashort X-ray pulses in a free-electron laser relies on
 high-density electron bunches with a precisely adjusted current and energy
 distribution.
To this end, robust and flexible electron bunch manipulation techniques are
 required that allow a high degree of control over the phase space density of
 the bunch.
This paper reports on the demonstration of ultrashort current spikes with
 femtosecond duration, created by compressing an electron bunch after a
 laser-induced energy modulation with linearly varying envelope.
This scheme is implemented at the free-electron laser FLASH, where the energy
 modulation is created early in the linear accelerator before the bunch is
 accelerated to its final energy.
Formation of the spikes is observed in measurements of the longitudinal
 phase space density.
It is demonstrated that, in conjunction with conventional compression
 techniques, this laser-based scheme allows to create two spikes
 with variable temporal separation.
Therefore, the demonstrated compression scheme shows great potential of
 enabling flexible ultrashort single- and double-pulse operation modes of
 free-electron lasers.
\end{abstract}

\maketitle

Free-electron lasers (FELs) operating in the soft to hard X-ray region 
\cite{Ackermann2007,Emma2010,Allaria2012,Kang2017,Decking2020,Prat:2020}
 play an important role in the research of ultrafast dynamics of atoms,
 molecules and condensed matter as they routinely produce high-brightness
 photon pulses with durations of a few tens of femtoseconds.
Enhancing the temporal resolution of these facilities even further requires
 the generation and control of FEL pulses on time scales of a few or even
 sub-femtoseconds.
To this end, several advanced schemes to shape the electron bunches have been
 explored allowing to control which parts of the bunch support the FEL process
 \cite{Huang2017,Schneidmiller2023,Prat2023,Marinelli2017,Marinelli2016,%
 Zhang2020,Duris2021,Amstutz2024IPAC,Li2024}.
By limiting the lasing region to only a fraction of the total bunch length,
 ultrashort FEL pulses can be produced.

Suitable bunch shaping has been achieved by either creating a short current
 spike or spoiling electron parameters crucial for the FEL process except for a
 short region within the bunch.
For instance, a short leading spike can be created by imprinting
 a nonlinear variation of the energy along the bunch via the radiofrequency
 (RF) cavities of the linear accelerator (linac) followed by longitudinal
 dispersion \cite{Huang2017,Schneidmiller2023,Prat2023}.
On the other hand, transverse dispersion in combination with beam-spoiling
 masks has been utilized for bunch shaping at low-repetition-rate FELs
 \cite{Marinelli2017}.
Various advanced bunch shaping techniques based on optical lasers
 \cite{Marinelli2016,Zhang2020,Duris2021,Amstutz2024IPAC,Li2024} have
 demonstrated the feasibility of generating ultrashort FEL pulses.
Laser-driven schemes have the advantage that they can be applied to a
 selectable subset of the electron bunches which enables flexible operation
 schemes at FELs with multiple user beamlines \cite{Li2024}.

This paper reports on the demonstration of a laser-based
 compression scheme based on energy modulation with a
 linearly varying amplitude as proposed in Ref.~\cite{Tanaka2019}.

An electron bunch, or a part thereof, that exhibits a correlation in
 longitudinal phase space (``energy chirp''), spanned by temporal position $t$
 and deviation $\Delta E \equiv E-E_0$ from the central energy $E_0$, is
 compressed in a longitudinally dispersive section, typically a magnetic
 chicane, of appropriate strength.
RF cavities that operate at a wavelength much larger than the bunch
 can imprint a correlation that encompasses all electrons,
 resulting in the compression of the bunch as a whole.
On the other hand, a laser pulse co-propagating with the electron bunch in an
 undulator (``modulator'') creates a sinusoidal optical-scale energy
 modulation.
The resulting correlation of the electrons around the zero crossings leads to a
 periodic compression, which is the basis for many FEL
 seeding schemes, see Ref.~\cite{Hemsing2014} and references therein.

\begin{figure}
\includegraphics{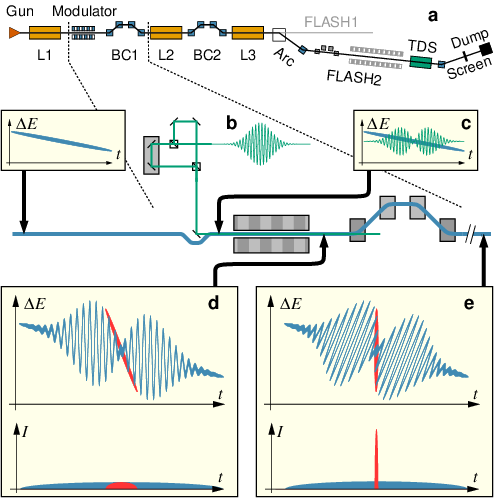}
\caption{%
Schematic layout of the FLASH accelerator and FEL beamline FLASH2 in the
 configuration used during the experiment (a), see Methods for details.
A laser pulse (green, wavelength not to scale) is
 sent through a two-path interferometer (b) and
 overlapped with an electron bunch (blue) in the modulator (c).
After energy modulation (d), a part of the longitudinal phase space density is
 linearly correlated (red).
Longitudinal dispersion in magnetic chicanes (BC1 and BC2) compresses the correlated
 electrons, forming a short current spike (e).
}
\label{fig:schematic}
\end{figure}

In an alternative laser-based compression scheme proposed in
 Ref.~\cite{Tanaka2019}, an optical-scale energy modulation with linearly
 varying amplitude  is imprinted onto the bunch.
A nearly linear correlation is created between electrons around minima and
 maxima of the modulation, see red regions in \figref{fig:schematic}.
The energy chirp added to these electrons is determined by the slope of the
 energy modulation envelope.
Downstream of the modulator, a magnetic chicane compresses the linearly
 correlated electrons and an ultrashort current spike is formed.
A suitable laser pulse is created from a Gaussian pulse in a
 two-path interferometer with a transit-time difference close to the pulse
 duration.
This compression scheme was implemented at the soft X-ray FEL user facility
 FLASH \cite{Ackermann2007} in Hamburg, Germany, where a two-path interferometer
 was added to the laser beamline of an existing modulator setup, which is used
 as a laser heater in standard operation \cite{Gerth2021IPAC,Gerth2023IPAC}, and
 the existing magnetic chicanes (BC1 and BC2) used for conventional RF-based
 bunch compression were utilized to form the spike.

\section*{Laser-based bunch shaping}
A two-path interferometer splits an incoming Gaussian laser pulse with peak
 electric field $\Ef_0$ and introduces a temporal separation between the
 partial pulses before they are recombined.
If the separation $\delta_t$ from their common center is not much larger than
 their root mean square (rms) electric field duration $\sigma_t$, the pulses
 overlap partially and the electric field in the overlap region is
 affected by interference effects.
In case of destructive interference, the resulting field envelope exhibits a
 nearly linear behavior in a region around the common center, as depicted in
 \figref[a]{fig:envelope} and \figref[b]{fig:envelope}.
Figure~\fref[c]{fig:envelope} shows the value of the first three derivatives of
 the envelope at the pulse center as a function of the relative pulse separation
 $\delta_t/\sigma_t$ (see Methods).
While the slope (red) determines the linear chirp of the correlated electrons,
 the third derivative (blue) is a measure for the nonlinearity of the envelope.
Due to the symmetry of the envelope function, the second
 derivative (orange) and all further even derivatives (not shown) vanish
 for all $\delta_t$.
If $\delta_t=\sigma_t$, the slope reaches its maximum absolute value of
 $\e^{-\frac12}\,\Ef_0/\sigma_t$.
On the other hand, also the third-order derivative is close to its maximum.
Nonlinear contributions limit the extent of the region in which the modulation
 is sufficiently linear for the compression scheme.
At a pulse separation of $\delta_t = \sqrt{3}\sigma_t$ the third derivative
 vanishes (\figref[c]{fig:envelope}) so that the nonlinear contributions are
 minimized.
Here, the relative deviation between
 the tangent (red in \figref[b]{fig:envelope}) and the envelope function
 (green) exceeds $5\%$ at $|t| \approx \sigma_t$, whereas for
 $\delta_t=\sigma_t$ this occurs at $|t| \approx 0.4 \sigma_t$, see
 \figref[a]{fig:envelope}.
The slope, however, is diminished to $\sqrt{3} \e^{-\frac32}\Ef_0/\sigma$,
 which is $64\%$ of the maximum value.
Pulse separations much larger than $\sqrt{3}\sigma_t$ are hardly suitable for
 the compression scheme, as the resulting slope rapidly approaches zero.
During the experiment reported here, the laser pulse separation was set close
 to $\delta_t=\sigma_t$ in order to maximize the energy chirp of the
 linearly correlated electrons.
\newcommand{\hindtheo}{\SI{-9.7(2.0)}{keV/ps}}
A laser-induced chirp of $\rmd E / \rmd t = \hindtheo$ was calculated from the
 laser and modulator parameters (see Methods).

\begin{figure}
\includegraphics{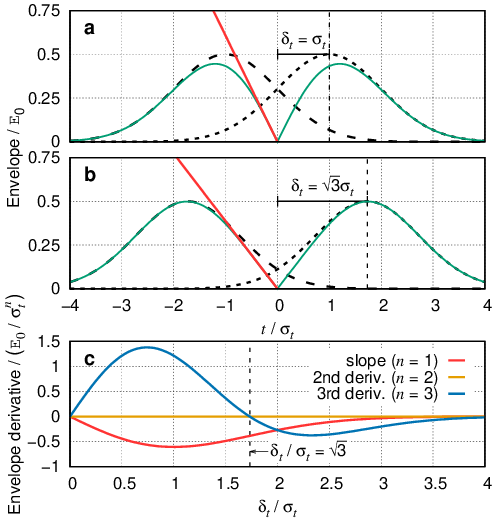}
\caption{%
Field envelopes of two individual Gaussian pulses (dashed black), their
combined envelope (green) in case of destructive interference, and tangent
(red) to the combined envelope at the pulse center ($t\to -0$).
Depicted are the pulse separations for maximum absolute slope of the tangent
 (a) and best linearity (b),
 as well as the time derivatives of the combined envelope at the center ($t\to -0$)
 under variation of the relative pulse separation $\delta_t/\sigma_t$ (c).
}
\label{fig:envelope}
\end{figure}

\begin{figure*}
\includegraphics{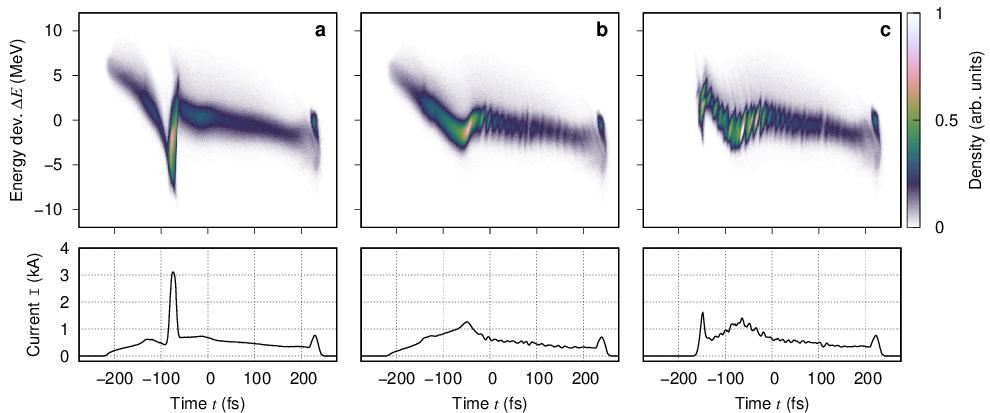}
\caption{%
Longitudinal phase space density (top) and current profile (bottom) measured
 downstream of the FLASH2 undulator beamline for three cases:
 both interferometer arms open (a),
 one interferometer arm blocked (b),
 laser blocked completely (c).
}
\label{fig:onoff}
\end{figure*}

\section*{Longitudinal phase space measurements}
A transverse-deflecting structure (TDS) setup located downstream of the FLASH2
 undulators (see \figref{fig:schematic}) allows direct measurements of the
 longitudinal phase space density, see Methods.
Figure \fref{fig:onoff} shows longitudinal phase space measurements
 (top) and deduced current profiles (bottom)
 for three different situations, illustrating the effect of laser-induced
 compression (Table~\ref{tab:linac} column a in Methods).
With both interferometer arms unblocked (\figref[a]{fig:onoff}), a pronounced
 spike of over $\SI{3}{kA}$ is created in the current profile on top of a
 pedestal of about $\SI{0.6}{kA}$.
Around the spike, a strong energy chirp is apparent in the
 phase space density,
 which can be attributed to energy kicks from longitudinal space charge and
 coherent synchrotron radiation in BC2 as well as the extraction arc \cite{Saldin1997}.
Blocking one interferometer arm (\figref[b]{fig:onoff})
 causes the current spike and energy chirp to vanish. 
Characteristic inhomogeneities resulting from the microbunching
 instability \cite{Saldin2002,Ratner2015}
 appear ahead of the former spike position,
 which shows that microbunching in this region was previously suppressed by
 the heating effect of the now blocked part of the laser pulse.
When the laser is blocked entirely (\figref[c]{fig:onoff}), microbunching
 structures appear behind the former spike position ($t<-\qty{80}{fs}$) as
 well.

Overall, these results verify that the observed
 spike is indeed created by the compression effect
 resulting from an energy modulation with linearly varying amplitude 
 induced by two destructively interfering laser pulses.

\subsection*{Scan of the RF-induced chirp}
Phase-space measurements have been recorded over a range of RF chirp
 settings for the first linac section L1 (see Table~\ref{tab:linac} column b in Methods).
Exemplary measurements for three RF chirp values are shown in
Figs.~\fref[a]{fig:chirpscan} to \fref[c]{fig:chirpscan} together with the
corresponding current profiles.
By fitting a Gaussian function on top of a linear background to the current
 profile in a region around the spike, its
 rms duration (\figref[d]{fig:chirpscan}) and
 peak current (\figref[e]{fig:chirpscan}) were determined.
 
In the sign convention used here, a more negative induced chirp corresponds to
 a stronger RF compression, until the bunch reaches its minimum duration
 (``full compression'') and is decompressed again for even lower chirps
 (``over-compression'').
For chirps below $\qty{-213}{keV/ps}$ the measured rms spike duration first
 decreases nearly linearly and then transitions into a flat region spanning
 $\qtyrange{-222}{-227}{keV/ps}$.
Beyond $\qty{-229}{keV/ps}$, the entire bunch comes close to
 full compression, and the laser-induced spike formation cannot be clearly
 distinguished from features stemming from the strongly RF-compressed
 bunch.
 
The temporal resolution of the TDS setup depends not only on parameters of the
 accelerator and the instrument but also on the beamsize along the temporal
 axis of the TDS screen.
As the beamsize varies along the temporal position in the bunch, the
 resolution of the TDS measurement is not constant.
In the flat region, the mean measured rms spike duration is approximately
 $\qty{5.2}{fs}$, which determines an upper limit for both the temporal
 resolution at the position of the spike and the actual spike duration.

\begin{figure}
\includegraphics{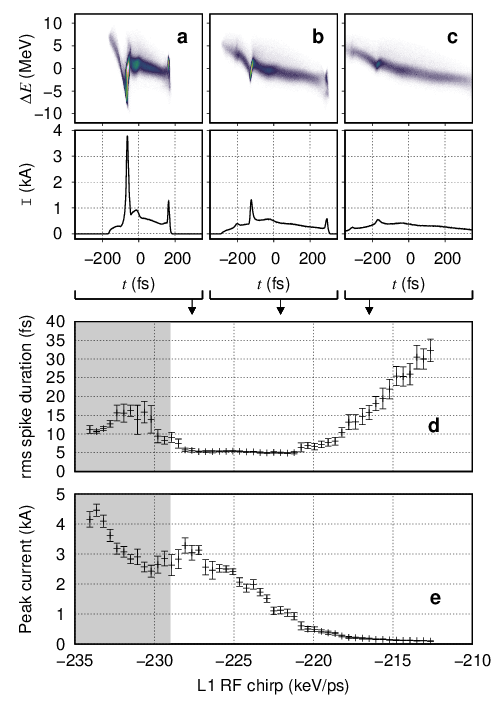}
\caption{
Examplary phase space densities and current profiles observed at the RF-induced
 chirp created in L1 indicated by the arrows (a--c) and measurements of the rms
 duration (d) and peak current (e) of the spike as a function of the chirp.
In the shaded region, the entire bunch comes close to or exceeds full
 compression.
}
\label{fig:chirpscan}
\end{figure}

\begin{figure}
\includegraphics{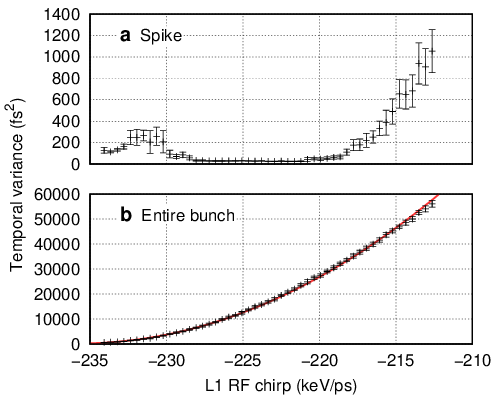}
\caption{
Temporal variance of the longitudinal charge density of
 the current spike created by the laser-induced compression (a)
 and the entire bunch (b) measured as a function of the energy chirp induced in
 linac L1 together with a quadratic fit (red).
}
\label{fig:chirpscan-variance}
\end{figure}

Figure~\ref{fig:chirpscan-variance} shows the temporal variance of the 
 entire bunch and the laser-induced spike determined from the same set of
 phase space measurements as presented in \figref{fig:chirpscan}.
The variance of the entire bunch in \figref[b]{fig:chirpscan-variance}
 exhibits a quadratic behavior as can be seen from the fit (red),
 which is compatible with linear beam dynamics theory (see Methods)
 and indicates that the bunch is fully compressed when a chirp of
 $h_b \equiv \SI{-235.7}{keV/ps}$ is induced in L1.
However, similar to the rms width of the spike (\figref[d]{fig:chirpscan}), its
 variance shown in \figref[a]{fig:chirpscan-variance} exhibits
 a pronounced flat region, which indicates that it cannot be attributed to the
 finite resolution of the measurement setup only:
Any resolution $\sigma_\text{res}$ manifests itself in the observed
 variance $\sigma_\text{obs}^2$ via an addition to the actual value
 $\sigma_\text{obs}^2 = \sigma_\text{act}^2 + \sigma_\text{res}^2$,
 which would result in an offset of the variance parabola, but not in a
 flattening of the curve.
A possible explanation could be related to the space-charge-induced positive
 energy chirp, which increases with compression.
In conjunction with spurious negative longitudinal dispersion in the FLASH2
 extraction arc (see \figref[a]{fig:schematic}), which can occur in case of
 optics mismatch \cite{Scholz2013Thesis}, this potentially leads to the
 spike being strongly compressed for a wider range of RF chirps.
Although the flat region (\figref[a]{fig:chirpscan-variance}) obscures the
 exact chirp at which the minimum of the temporal variance of the spike occurs,
 it must be within approximately $h_a \equiv \SI{-224.5(2.5)}{keV/ps}$.
As shown by Eq.~\eqref{eq:chirpdiff} in the Methods section, the difference of
 the chirps at the minima of these two curves $h_b-h_a$ is equal to the initial
 chirp of the linearly correlated electrons from the energy modulation.
Therefore, a laser-induced chirp of $\SI{-11.2(2.5)}{keV/ps}$ can be deduced,
 which is consistent with the value $\hindtheo$ calculated from
 parameters of the energy modulation (see Methods).
The good agreement between measurement and theory shows the viability of the
 model to describe the dynamics of the spike formation process.

\subsection*{Generation of two spikes with variable separation}
A decisive advantage of the laser-based compression scheme is that the
 position of the induced current spike within the bunch is determined by the
 relative timing of the modulating laser pulse and the electron bunch at the
 modulator, which can be controlled with high accuracy by an optical-fiber-based
 synchronization system \cite{Schulz2013IBIC}.
As shown in the following, the ability to control the spike position via
 the laser timing allows to implement a double-spike scheme with variable
 temporal separation, where a second spike is created by means of nonlinear
 RF-based bunch compression at the head of the bunch,
 c.f.~\cite{Huang2017,Schneidmiller2023,Prat2023}.
To produce this second spike, the nonlinear energy chirp imprinted by linac
 L1, which contains a third-harmonic accelerating module, was optimized.

Exemplary phase space measurements for three settings of laser timing
 $\Delta t_\text{L}$
 relative to the region forming the spike at the head
 are presented in
 Figs.~\fref[a]{fig:timingscan} to \fref[c]{fig:timingscan}
 together with the corresponding current profiles, each featuring two current
 spikes.
The spike at $t\approx\qty{110}{fs}$ was generated by nonlinear RF-based
 compression, whereas the other spike was created by the
 laser-induced compression scheme.
The temporal separation between the two spikes $\Delta t_\text{S}$ 
 as a function of the laser timing $\Delta t_\text{L}$
 is shown in \figref[d]{fig:timingscan}.
For each laser timing, the spike separation of 32 current profiles was
 determined from fits of two Gaussian functions with linear background.
As the laser-induced chirp is imprinted onto the bunch upstream of
 the first chicane BC1, the spike separation is subject to
 bunch compression and $\Delta t_\text{L}/\Delta t_\text{S}$ reflects the
 compression factor, which was determined to be $62.5\pm1.5$ from a linear fit
 to the data.
The theoretical bunch compression factor calculated from the chicane and RF
 parameters (see Table~\ref{tab:linac} column c in Methods) is $47.4$, assuming
 linear dynamics and an initial chirp of $\qty{21.1}{keV/ps}$, as determined
 from the RF chirp scan of the entire bunch
 (\figref[b]{fig:chirpscan-variance}) by applying Eqs.~\eqref{eq:Eqqi},
 \eqref{eq:Eqpi}, and \eqref{eq:defhpsi} (see Methods).
Utilizing a nonlinear RF chirp to create the second spike renders the linear
 approximation underlying the theoretical compression factor imprecise, which
 can explain the deviation from the measurement.

When the laser timing value is larger than the rms laser pulse duration
 of approximately \qty{4.8}{ps}, the bunch region that forms the spike
 at the head is not strongly affected by the induced energy modulation so that
 leading spike is largely unperturbed (Figs.~\fref[b]{fig:timingscan} and
 \fref[c]{fig:timingscan}).
At smaller timing values, the laser pulse overlaps with this region and
 increases its energy spread, which leads to a reduction of the peak current in
 the RF-induced spike (\figref[a]{fig:timingscan}).
Also, the peak current of the laser-induced spike is reduced when created
 close to the head of the bunch.
In this case, the laser-induced structure is not fully compressed, as the shape
 of phase space density is dominated by the nonlinear RF-induced chirp
 that results in the formation of the leading spike.
Nevertheless, at $\Delta t_\text{L} \approx \qty{2}{ps}$ still two distinct
 spikes are created with a separation of around \qty{25}{fs}
 (\figref[a]{fig:timingscan}).
At smaller separations, they cease to be distinguishable, partially due to the
 limited temporal resolution of the measurement.

At the modulator, the arrival time jitter of both electron bunch and laser
 pulse is of the order of \qty{100}{fs}, so that their relative jitter can be
 estimated to be approximately \qty{150}{fs}.
In comparison, due to the RF-based compression, the shot-to-shot jitter of the
 spike separation at the end of the linac is reduced by the compression factor.
At a spike separation of \qty{25}{fs}, the rms jitter is approximately
 \qty{4}{fs} and increases nearly linearly to about \qty{7}{fs} at a separation
 of \qty{180}{fs}, see \figref[d]{fig:timingscan}.
While the jitter at small separation is dominated by the relative arrival-time
 jitter at the modulator, the proportional increase can be attributed to the
 shot-to-shot variation of the compression factor resulting from RF amplitude
 fluctuations.
 
\begin{figure}
\includegraphics{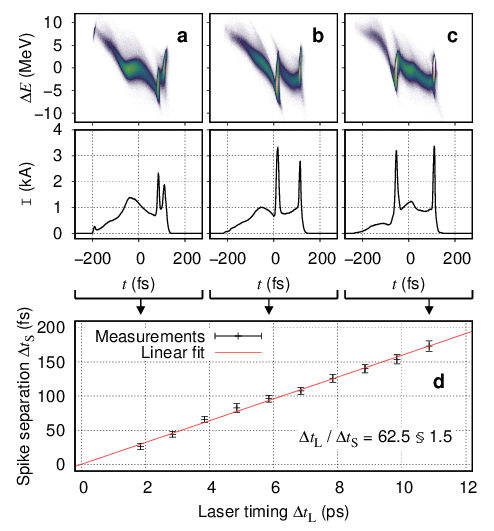}
\caption{
Three examples of the phase space density and current profile observed at the
 laser timing indicated by the arrows (a--c) and separation between the two
 spikes measured as a function of the laser timing (d) with the arbitrary
 reference value of the laser timing chosen such that the linear fit (red)
 passes through the origin, implying that
 at $\Delta t_\mathrm{L}=0$ the laser- and RF-induced spikes overlap.
}
\label{fig:timingscan}
\end{figure}

\section*{Discussion}
To the best of the authors' knowledge, the experimental results reported here
 constitute the first demonstration of the novel bunch compression scheme
 proposed in Ref.~\cite{Tanaka2019}.
Measurements of the longitudinal phase space density show that it enables the
 generation of a current spike with an rms duration shorter than \SI{5.2}{fs}.
The spike duration measured as a function of the RF chirp is compatible with a
 linear model of the spike formation dynamics --
 apart from a flat region (see Figs.~\fref[d]{fig:chirpscan} and
 \fref[a]{fig:chirpscan-variance}), which is presumably related to space charge
 effects and calls for further investigation.
The results show the great potential of this scheme for generating few- or
 possibly sub-femtosecond FEL pulses in the soft to hard X-ray regime. 
It was demonstrated that the scheme allows to control the
 spike position within the bunch via the timing of the laser pulse.
In combination with nonlinear RF compression, this represents a viable
 strategy for the implementation of a flexible FEL double-pulse scheme.

While Ref.~\cite{Tanaka2019} proposed to install a dedicated
 modulator and magnetic chicane at full beam energy after the main
 linac, the implementation reported here utilizes an existing modulator setup
 at low beam energy, which is commonly installed at X-ray FEL facilities
 as a laser heater in order to suppress microbunching instabilities.
An advantage of early modulation is that the relative arrival time jitter
 between electron bunch and laser pulse at the modulator does not directly
 translate into the spike-separation jitter in the double-spike scheme
 described above, but is reduced due to bunch compression in the subsequent
 linac.
On the other hand, the intrinsic synchronization between laser and
 current spike is partly spoilt by the jitter in the linac.
This is a disadvantage for pump-probe experiments
 with a pump pulse linked to the modulator laser timing.
Adding the interferometer to an existing laser heater setup facilitates
 swift and cost-efficient implementation.
Facilities with multiple FEL beamlines would benefit from adding a
 dedicated laser system, which allows to apply the compression scheme to only a
 subset of electron bunches and independently of the laser heater, so that
 bunches can be tailored for the individual beamlines, c.f.~\cite{Li2024}.
At FLASH, this would allow to send long laser-heated bunches to the FLASH1 beamline,
 where seeding schemes at full energy are being implemented \cite{Schaper2021},
 while simultaneously laser-induced current spikes are used in
 the FLASH2 beamline to produce ultrashort FEL pulses.

\section*{Methods}

\subsection*{Experimental setup}
At FLASH, electron bunches are produced in an RF photocathode gun
 with a central energy of $E_0=\SI{5.6}{MeV}$
 and an rms duration of approximately $\SI{5}{ps}$.
As depicted in \figref[a]{fig:schematic}, the FLASH linac consists of 
 three superconducting acceleration sections L1 to L3, which accelerate the
 bunches to $\SI{143.0}{MeV}$, $\SI{560.5}{MeV}$, and up to $\SI{1350}{MeV}$),
 respectively.
Bunches can be distributed among the two FEL beamlines FLASH1 and
 FLASH2 \cite{Faatz2016}, where the latter is connected to the linac via an
 extraction arc \cite{Scholz2011FEL}.
In this study, the final beam energy was $\SI{1213.5}{MeV}$.
The longitudinal dispersion $M_{56}$ of the magnetic bunch compression chicanes
 BC1 and BC2 was set to $\qty{157.7}{mm}$ and $\qty{72.8}{mm}$, respectively.

Downstream of the FLASH2 undulators, an X-band
 transverse-deflecting structure (TDS) \cite{Marchetti2021,Vogt2022FEL}
 in combination with the transverse dispersion of the dump section allows to
 directly measure the longitudinal phase space density of the electron bunches.
The transverse momentum imparted to an electron depends linearly on its
 arrival time at the TDS, which further downstream results in a strong correlation
 between the longitudinal and transverse position.
In the perpendicular transverse direction, dispersion correlates
 an electron's energy with its position.
Due to these correlations, the charge density measured on a screen reveals the
 longitudinal phase space density of the bunch.

As the focus of the measurements is on the compression effect itself, the
 FLASH2 permanent-magnet undulators were opened to their maximum gap in order
 not to obscure the longitudinal phase space density with features created by
 the FEL process.
The magnetic chicane between the extraction arc and the FLASH2 undulators was
 turned off, see \figref[a]{fig:schematic}.

An RF chirp configuration that optimizes the laser-induced compression effect
 was initially estimated based on the expected chirp of the linearly correlated
 electrons and subsequently fine-tuned empirically via the chirp created by L1
 while observing the longitudinal phase space density.

The permanent-magnet modulator
 with period length $\lambda_\mathrm{u}=\SI{43}{mm}$,
 $N_\mathrm{u} = 11$ full periods,
 undulator parameter $K=1.37$,
 and modified undulator parameter
 (see Ref. \cite{SchmueserFEL})
 $\widehat{K} \equiv K \cdot JJ = 1.11$
 is located downstream of the linac section L1
 and is used as a laser heater in standard operation.
It is driven by laser pulses with
 a central wavelength of $\SI{532}{nm}$,
 rms field duration $\sigma_t = \SI{4.7(0.3)}{ps}$,
 pulse energy $U_\mathrm{L} = \SI{55(5)}{\micro J}$,
 and an average effective rms spot size of
 $\sigma_\perp = \SI{350(25)}{\micro m}$, which was estimated from screen
 images up- and downstream of the modulator with the indicated estimated
 systematic uncertainties~\cite{Gerth2023IPAC}.
With the on-axis laser intensity in the modulator
 \begin{equation}
 I_\mathrm{L} = \frac{P_\mathrm{L}}{2\pi \sigma_\perp^2} = \frac{\Ef_0^2}{2 Z_0},
 \end{equation}
 where the laser power is
 $P_\mathrm{L}= U_\mathrm{L} / \sqrt{\pi\sigma_t^2}$,
 the on-axis electric field amplitude $\Ef_0$ and the peak amplitude of the
 resulting energy modulation $A_E$ can be seen to be
 \begin{equation}
 \Ef_0 = \frac1{\sigma_\perp}\sqrt{\frac{P_\mathrm{L} \, Z_0}{\pi}}
 \quad \text{and} \quad
 A_E = \frac{e \Ef_0 \widehat{K} N_\mathrm{u} \lambda_\mathrm{u}}{2\gamma},
 \end{equation}
 respectively, with the free-space impedance
 $Z_0/\pi \approx \SI{120}{\Omega}$,
 Lorentz factor $\gamma$,
 and elementary charge $e$
 \cite{Griffiths1999,SchmueserFEL}.
The above values and error bounds yield $A_E = \SI{75(16)}{keV}$,
 which for $\delta_t=\sigma_t$ results in a laser-induced energy chirp of
 $\e^{-\frac12} A_E/\sigma_t = \hindtheo$, as derived in the next section.

\begin{table}[h!]
\caption{%
RF parameters (central energy $E_0$ and added chirp $h$)
and average bunch charge
during the measurements presented in
\figref{fig:onoff} (a),
\figref{fig:chirpscan} (b),
and 
\figref{fig:timingscan} (c).
}
\label{tab:linac}
\begin{tabular*}{\linewidth}{c@{\extracolsep{\fill}}cccc}
\hline \hline
	%   empty
    &   $E_0$ (MeV)
	&   \multicolumn{3}{c}{$h$ (\unit{keV \per \pico\second})}
	\\
	\cline{3-5}
   &                    & a     & b              & c     \\
\hline
L1 & $\phantom{1}143.0$ & $-227$  & $-213$ to $-234$ & $-223 $ \\
L2 & $\phantom{1}560.5$ & $-1278$ & $-1278$          & $-1329$ \\
L3 &           $1213.5$ & $0$     & $0$              & $0$     \\
\hline
charge      &  & \SI{260}{pC}     & \SI{225}{pC}           & \SI{270}{pC} \\
\hline \hline
\end{tabular*}
\end{table}

\subsection*{Calculation of the laser envelope}
An expression for the envelope of two interfering Gaussian laser pulses is
 derived, which is defined as the absolute value of the analytic signal of the
 combined pulse, cf.~\cite{Osgood}.
Generally, the analytic signal $\A f(t)$ of a function $f(t)$
 is constructed by setting the negative-frequency Fourier components to zero,
 doubling the strictly positive-frequency components, and keeping the DC
 component
	\begin{equation}
	\F \A f(\omega)
	\equiv
	%\F^{-1}/
	%2 \, \Theta_{\frac12}(\omega) \, \F f(\omega) \\
	%&=
	%\F^{-1}
	\begin{cases}
		0, & \omega < 0 \\
		\F f(0), & \omega = 0 \\
		2 \F f(\omega), & \omega > 0
	\end{cases},
	\end{equation}
where $\F$ denotes the Fourier transform operator with
	$
	\F g(\omega)
	\equiv
	\frac{1}{\sqrt{2\pi}}
	\int_\R \, \e^{-\I \omega t} \, g(t) \, \rmd t.
	$

Consider $\omega_0 \in \R^+$, $A\!\!: \R \to \R$ and
 $f\!\!: \R \to \R$
 with  $f(t) = A(t) \cos(\omega_0 t)$.
It can be seen that
	$
	2 \F f(\omega) =
	\F A(\omega - \omega_0) + \F A(\omega + \omega_0)
	$.
Introducing the residual term
	\begin{equation}
	\label{eq:Feps}
	\F \varepsilon(\omega)
	\equiv
	\begin{cases}
		- \F A(\omega - \omega_0), & \omega < 0 \\
		0 & \omega = 0 \\
		+ \F A(\omega + \omega_0), & \omega > 0
	\end{cases},
	\end{equation}
 allows to write
	$
	\F \A f (\omega) = \F A(\omega - \omega_0) + \F \varepsilon(\omega)
	$
 and therefore
	\begin{equation}
	\A f(t) = \e^{\I \omega_0 t} A(t) + \varepsilon(t).
	\end{equation}
As $A(t)$ is a real function, it follows that
 $\F A(\omega) = \F A(-\omega)^{*}$.
Using this relation, it becomes apparent from Eq.~\eqref{eq:Feps} that
 $\F \varepsilon(\omega) = -\F \varepsilon(-\omega)^{*}$,
 which implies that $\varepsilon(t)$ is an imaginary-valued function.
Hence, the original function $f(t)$ is recovered by taking
 the real part of the analytic signal $f(t) = \mathrm{Re}\{\A f(t)\}$.

\begin{figure}
\includegraphics{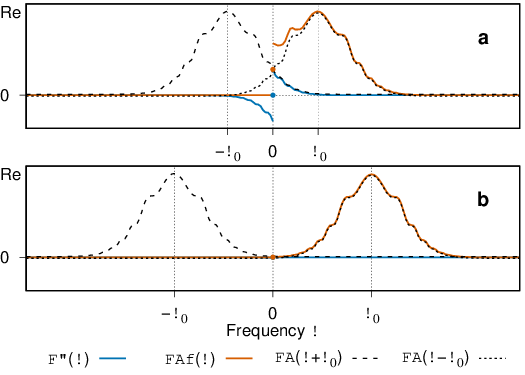}
\caption{
Illustration of the approximation in Eq.~\eqref{eq:highfreqapprox}.
(a) If the bandwidth of $A(t)$ is larger than $\omega_0$,
 then the Fourier spectrum of the analytic signal $\F \A f(\omega)$ differs
 significantly from $\F A(\omega-\omega_0)$ at small absolute frequencies, as
 quantified by $\F \varepsilon(\omega)$.
(b) If the spectral width is much smaller than $\omega_0$, this
 difference vanishes.
}
\label{fig:envtheo}
\end{figure}

If the Fourier spectrum of $A(t)$ is well localized and its bandwidth is
 much smaller than $\omega_0$, then the contribution of $\varepsilon(\omega)$
 to the analytic signal of $f$ is small and might be neglected,
 as illustrated in \figref{fig:envtheo}.
In this high-frequency case, the analytic signal of $f$
is well approximated  by
	\begin{equation}
	\label{eq:highfreqapprox}
	\A f(t) \approx \e^{\I \omega_0 t} A(t)
	\end{equation}
 and therefore $|\A f(t)| \approx |A(t)|$.

\newcommand{\Ein}{f_\mathrm{in}}
\newcommand{\Eout}{f_\mathrm{out}}

Consider
	$
	\Ein(t) \equiv \Ef_0 \, \xi_\sigma(t) \, \cos(\omega_0 t)
	$
 with $\Ef_0,\sigma,\omega_0 \in \R$,
 describing the pulse entering the interferometer,
 where $\omega_0 \gg \sigma^{-1}$ and
	$
	\xi_\sigma(t) \equiv \e^{-\frac12(t/\sigma)^2}
	$.
From $\F \xi_\sigma \propto \xi_{\sigma^{-1}}$,
 it follows that
 $|\F \xi_\sigma(\omega) | \ll 1 \quad \forall \omega > \omega_0$,
 so that in the following the above approximation of
 the analytic signal of $\Ein$ can be taken as exact,
	$
	\A \Ein(t) = \Ef_0 \e^{\I \omega_0 t} \xi_\sigma(t).
	$
Let
	$$
	\Eout(t) \equiv \frac12 \left[\Ein(t+\delta) + \Ein(t-\delta)\right]
	$$
 with $\delta \in \R^+$, i.e., half the superposition
 of two copies of $\Ein(t)$ shifted relative to each other
 by $2\delta$, which describes the pulse after the interferometer.
Due to the linearity of the $\A$ operator,
 the analytic signal of $\Eout(t)$ is -- within the
 high-frequency approximation -- given by
	\begin{align}
	\A \Eout(t)
	& =
	\tfrac{1}{2}
	[\A\Ein(t+\delta) + \A\Ein(t-\delta)] \\
	& =
	\tfrac{1}{2}
	\left[
	\e^{z(t)}
	+\e^{-z(t)}
	\right] \,
	\xi_\sigma(\delta) \,
	\A\Ein(t),
	\end{align}
 where $z(t) \equiv \I \omega_0 \delta -\tfrac{\delta}{\sigma^2} t$.
It can be seen that
	\begin{align}
	|\A\Eout(t)|^2
	&= \A\Eout(t) \, \A\Eout(t)^* \\
	&= \tfrac{1}{4}|\xi_\sigma(\delta) \, \A\Ein(t)|^2
	  \sum%_{\pm,\pm}
	  \e^{\pm z(t) \pm z(t)^*}
	,
	\end{align}
 where the sum runs over all four combinations of the signs
 of $z(t)$ and $z(t)^*$.
Employing well-known trigonometric identities, this results in
	\begin{equation}
	\label{Sat}
	|\A\Eout(t)| =
	\tfrac{1}{\sqrt{2}}
	\Ef_0
	\xi_\sigma(\delta) 
	\xi_\sigma(t)
	\sqrt{
		  \cosh(\tfrac{2\delta}{\sigma^2}t)
		+ \cos(2 \omega_0 \delta)
	}.
	\end{equation}

Destructive interference occurs when
 $\delta = (m+\tfrac{1}{2})\tfrac{\pi}{\omega_0} \equiv \delta_-$
 with $m \in \Z$, which implies $\cos(2 \omega_0 \delta) = -1$.
By introducing the normalized pulse seperation
 $\nu \equiv \delta_- / \sigma$ and symbol
 $\check A(t) \equiv |\A\Eout(t)|_{\delta=\delta_-}$
 for the resulting pulse envelope, Eq.~\eqref{Sat} can be written as
	\begin{align}
	\check A(t)
	&=
	\frac{1}{\sqrt{2}}
	\Ef_0
	\sqrt{
	\e^{-\left(\frac{t^2}{\sigma^2} + \nu^2\right)}
	\left(
		\frac{
			\e^{ 2\nu\frac{t}{\sigma}} +
			\e^{-2\nu\frac{t}{\sigma}}
		}{2}
		-1
	\right)
	}
	\\&=
	\frac12
	\Ef_0
	\left|
	\e^{-\frac12\left(\frac{t}{\sigma} + \nu\right)^2}
	-
	\e^{-\frac12\left(\frac{t}{\sigma} - \nu\right)^2}
	\right|,
	\end{align}
which shows that, in the high-frequency approximation, the combined envelope of
 two destructively interfering Gaussian pulses is given by the absolute value of
 the difference of their individual envelopes.

Of particular importance is the behavior of the envelope at its center, $t=0$.
As $\check A(t)$ is the absolute value
 of an odd function,
 its even derivatives vanish at $t=0$,
 while its odd derivatives are undefined.
However, in the limit $t\to\pm0$ it can be seen that
\begin{equation}
	\lim_{t\rightarrow \pm 0}
	\frac{\mathrm{d}^n}{\mathrm{d}t^n}
	\check A(t)
	=
	\begin{cases}
	0, & \text{if $n$ is even} \\
	\mp
	\frac{\Ef_0}{\sigma^n}
	\frac{\rmd^n}{\rmd\nu^n}\e^{-\frac12\nu^2},
		& \text{if $n$ is odd}
	\end{cases}.
\end{equation}
Note that the odd derivatives for $\lim_{t\to\pm0}$ are equal to
those of a Gaussian envelope $\Ef_0 \xi_\sigma(t)$ at $t=\mp\nu\sigma$.

The maximum slope at the pulse center is attained when $\nu=1$, that is,
 when the pulses are separated by $2\sigma$, in which case
$
	\check A(t) = \frac{\Ef_0}{\sigma} \e^{-\frac12} |t| + O(t^3).
$

When $\nu=\sqrt{3}$, the third derivative of $\check A(t)$
vanishes in the central limit.
Remarkably, the pulse envelope is then linear around the center up to fifth
 order,
$
	\check A(t) = \frac{\Ef_0}{\sigma} \sqrt{3} \e^{-\frac32} |t| + O(t^5).
$

\subsection*{Theory of the RF chirp scan}
It is shown how the covariance matrix of the longitudinal
 phase space density of a bunch can be reconstructed from measurements
 of the bunch length as a function of an energy chirp applied upstream of the
 point of measurement.
Beam dynamics is described using the phase space coordinates $\vec z = (q,p)^\T$,
 where $q \equiv t-t_0(s)$ and $p \equiv E-E_0(s)$ are the deviations of the
 temporal position and electron energy from the reference values $t_0(s)$
 and $E_0(s)$ at the location $s$ along the beamline.
When the beam is ultrarelativistic, the longitudinal single-particle dynamics
 in an FEL linac is dominated by the effects of accelerating structures and
 magnetic chicanes, which are described using a linear model:
 an RF module acts via the map
 $\vec z \mapsto \mat K(h) \vec z$
 with the kick matrix
	\begin{equation}
		\mat K(h) \equiv
		\begin{pmatrix}
			1 & 0 \\
			h & 1
		\end{pmatrix},
	\end{equation}
 where the energy chirp $h$ is determined by the RF settings;
 magnetic chicanes act via the map
 $\vec z \mapsto \mat D(\beta) \vec z$
 with the drift matrix
	\begin{equation}
		\mat D(\beta) \equiv
		\begin{pmatrix}
			1 & \beta \\
			0 & 1
		\end{pmatrix},
	\end{equation}
 where the longitudinal dispersion $\beta$ is related
 to the usual matrix element $M_{56}$ of the chicane via
 $\beta = M_{56}/(c\,E_0)$.
A typical FEL linac has multiple bunch compression stages, each
 comprising an accelerating section followed by a magnetic chicane.
In the linear model, the total map of $N$ compression stages is given by
$\vec z \mapsto \mat M \vec z$,
 where
	\begin{align}
		\mat M
	&= \mat M(h_1, \beta_1, \dots, h_N,\beta_N) \\
	&= \mat D(\beta_N) \mat K(h_N) \, \cdots \, \mat D(\beta_1) \, \mat K(h_1) .
	\end{align}
$\mat M$ can be decomposed into the form
	\begin{equation}
		\label{MKSD}
		\mat M = \mat K(h^\dagger) \mat S(C^\dagger) \, \mat D(\beta^\dagger)
	\end{equation}
 with the symplectic scaling matrix
	\begin{equation}
		\mat S(x) \equiv
		\begin{pmatrix}
			x^{-1} & 0 \\
			0 & x
		\end{pmatrix},
	\end{equation}
 where closed-form solutions for $h^\dagger$, $C^\dagger$ and
 $\beta^\dagger$ in dependence of $h_1,\beta_1,\dots,h_N,\beta_N$ exist
 \cite{Amstutz2023Thesis}.
The quantities
 $h^\dagger$, $C^\dagger$ and $\beta^\dagger$
 describe the \emph{total} chirp, compression and longitudinal dispersion
 created by the beamline.

The covariance matrix of a longitudinal phase space density
 $\Psi\!: \R^2 \to \R$ is defined
 by
	\begin{equation}
		\mat \Sigma[\Psi] \equiv
		\begin{pmatrix}
			\E{q^2}_\Psi & \E{q p}_\Psi \\
			\E{q p}_\Psi & \E{p^2}_\Psi
		\end{pmatrix},
	\end{equation}
 with the expected value defined by
 $\E{f(\vec z)}_\Psi
 \equiv \int_{\R^2} f(\vec z-\vec \mu_{\vec z}) \Psi(\vec z) \, \rmd \vec z$,
 where $\vec \mu_z \equiv \int_{\R^2} \vec z \Psi(\vec z) \, \rmd \vec z$.
Under a linear map defined by a symplectic matrix $\mat M$, an initial
 phase space density $\Psi_i$ evolves into the final one
 $\Psi_f(\vec z)
 = \Psi_i(\mat M^{-1} \vec z) \equiv (\Psi_i \circ \mat M^{-1})(z)$,
 whose covariance matrix can be seen to be
	$
	\mat \Sigma[\Psi_f] = \mat{M} \, \mat \Sigma[\Psi_i] \, \mat{M}^\T.
	$
Plugging in Eq.~\eqref{MKSD} for $\mat M$  yields
	\begin{gather}
	\mat \Sigma[\Psi_f] =
	\mat K(h^\dagger) \, \cdot \\
	\nonumber
	\begin{pmatrix}
		\frac{\beta^{\dagger 2} \Eppi
		+ 2 \beta^\dagger \Eqpi
		+ \Eqqi}{C^{\dagger 2}} &
		\beta^\dagger \Eppi + \Eqpi \\
		\beta^\dagger \Eppi + \Eqpi &
		C^{\dagger 2} \Eppi
	\end{pmatrix}
	\mat K(h^\dagger)^\T.
	\end{gather}
In particular, as $\E{q^2}_\Psi$ is invariant under kick maps, the above shows
 that
	\begin{equation}
	\label{q2dag}
	\E{q^2}_{\Psi_f} = 
	%\E{q^2}_{\Psi_f \circ \mat K(h^\dagger)^{-1}} = 
	\frac{\beta^{\dagger 2} \Eppi
	+ 2 \beta^\dagger \Eqpi + \Eqqi}{C^{\dagger 2}}.
	\end{equation}

In the FLASH linac, there are two bunch compression stages.
For a two-stage setup, it can be shown that
$\beta^\dagger = C^\dagger \, \bar \beta$
 and
$C^\dagger  =   \left[(h_1 + \bar h)\bar \beta\right]^{-1}$,
 where
$\bar \beta \equiv \beta_1 + \beta_2 + \beta_1 \beta_2 h_2$
 and
$\bar h \equiv (1+h_2 \beta_2) / \bar \beta$.
Equation~\eqref{q2dag} can then be written as
\begin{equation}
\label{EqqPsif}
\E{q^2}_{\Psi_f} = a_2 \left( h_1 - a_1 \right)^2 + a_0
\end{equation}
 with
	$a_2 \equiv \bar \beta^2 \Eqqi$,
	$a_1 \equiv -\left(\bar h + {\Eqpi}/{\Eqqi}\right)$,
	and
	$a_0 \equiv \bar \beta^2 \left( \Eppi - {\Eqpi^2}/{\Eqqi} \right)$.
Solving for the elements of $\Sigma[\Psi_i]$ yields
\begin{align}
	\label{eq:Eqqi}
	\Eqqi & = \frac{a_2}{\bar \beta^2}, \\
	\label{eq:Eqpi}
	\Eqpi & = - \left( a_1 + \bar h \right) \Eqqi, \\
	\label{eq:Eppi}
	\Eppi & = \frac{a_0}{\bar \beta^2} + \frac{\Eqpi^2}{\Eqqi}.
\end{align}
Within this model, the final temporal variance $\E{q^2}_{\Psi_f}$
 of a bunch depends quadratically on the RF-induced chirp $h_1$
 as shown in Eq.~\eqref{EqqPsif}.
The quantities $a_2$, $a_1$, and $a_0$ can therefore
 be determined experimentally by scanning $h_1$ and fitting a quadratic
 function to measurements of the resulting temporal variance.
If the machine parameters $\bar h$ and $\bar \beta$ are known,
 then the components of the covariance matrix can be explicitly reconstructed
 via Eqs.~\eqref{eq:Eqqi} to \eqref{eq:Eppi}.

For two chirp scans of two independent initial phase space densities
 $\Psi_{A}$ and $\Psi_{B}$,
 conducted with the same machine parameters $\bar h$ and $\bar \beta$, it
 follows from Eqs.~\eqref{eq:Eqqi} and \eqref{eq:Eqpi}, respectively, that the
 ratio of the temporal variances
 \begin{equation}
 	\label{eq:varratio}
	\frac{\EqqA}{\EqqB} = \frac{a_{2,A}}{a_{2,B}}
 \end{equation}
 and the difference of the energy chirps
 of the initial phase space densities
 \begin{equation}
	\label{eq:chirpdiff}
	h[\Psi_A]-h[\Psi_B] = a_{1,B} - a_{1,A}
 \end{equation}
 can be determined without knowledge of the machine parameters,
 where $h[\Psi]$ is defined as the energy chirp of $\Psi$,
 \begin{equation}
 	\label{eq:defhpsi}
 	h[\Psi] \equiv \frac{\Eqp}{\Eqq}.
 \end{equation}
Equation~\eqref{eq:chirpdiff} is used to determine the chirp imprinted by the
 laser pulse from the RF chirp scans presented in \figref{fig:chirpscan}.

The modulator used at FLASH for the laser-based compression scheme is located
 between the first linac section L1 and first bunch compression chicane BC1.
As the energy kicks from the modulator commute with those from the linac
 section, the above method reconstructs the covariance matrix of the
 phase space density upstream of the first linac L1 section plus the effects of
 the laser heater.

Collective kicks upstream of BC2, where the bunch current is low, are neglected
 in this model.
It is assumed that downstream of BC2 the longitudinal dispersion
 is negligible, so that energy kicks from collective effects such as
 longitudinal space charge and coherent synchrotron radiation experienced by
 the bunch after BC2 do not influence the bunch length measured at the TDS.
Spurious dispersion created in the extraction arc could void this assumption
 and potentially lead to the flat region observed in the spike variance under
 variation of the chirp induced by L1, which is not reflected by this model.
Nevertheless, for the purpose of estimating the induced chirp via
 Eq.~\eqref{eq:chirpdiff}, it can be assumed that for vanishing dispersion the
 variance minimum of the spike lies in the flat region.
Compared to the spike, the influence of collective effects on the entire bunch
 is small, so that for small spurious dispersion this model is still adequate
 and the covariance matrix can be estimated via  Eqs.~\eqref{eq:Eqqi} to
 \eqref{eq:Eppi}.

\begin{acknowledgments}
The authors would like to thank all colleagues at FLASH who ensured smooth
 operation of the accelerator and all its subsystems during the beamtimes.
This work is supported by the
 German Federal Ministry of Education and Research
 under grant number 05K22PE1.
\end{acknowledgments}

\bibliography{flare}

\end{document}